\documentclass[twocolumn]{aastex631}

\usepackage[version=4]{mhchem}
\usepackage{comment}
\usepackage{multirow}
\usepackage{array}
\usepackage{xcolor}
\usepackage{color}
\usepackage{soul}

\def\gsim{~\rlap{$>$}{\lower 1.0ex\hbox{$\sim$}}}
\def\lsim{~\rlap{$<$}{\lower 1.0ex\hbox{$\sim$}}}

\shorttitle{Climates of Gl 514 b}
\shortauthors{Delgado Diaz et al.}
\graphicspath{{./}}
\begin{document}

\title{Climates of Gl 514 b}

\author[0000-0002-8928-3929]{H\'ector E. Delgado D\'iaz}
\affiliation{Department of Astronomy, University of Washington, Seattle, WA 98195-1580, USA} 
\affiliation{NASA Virtual Planetary Laboratory, USA}

\author{Rory Barnes}
\affiliation{Department of Astronomy, University of Washington, Seattle, WA 98195-1580, USA} 
\affiliation{NASA Virtual Planetary Laboratory, USA}

\author[0000-0001-9423-8121]{Russell Deitrick}
\affiliation{School of Earth and Ocean Sciences, University of Victoria, Victoria, BC, CA}

\author{Mario Damasso}
\affiliation{INAF - Osservatorio Astroﬁsico di Torino, Via Osservatorio 20, I10025 Pino Torinese, IT}

\author{Nathaniel Masumoto Brown}
\affiliation{Department of Astronomy, University of Washington, Seattle, WA 98195-1580, USA}

%% Note that the \and command from previous versions of AASTeX is now
%% depreciated in this version as it is no longer necessary. AASTeX 
%% automatically takes care of all commas and "and"s between authors names.

%% AASTeX 6.31 has the new \collaboration and \nocollaboration commands to
%% provide the collaboration status of a group of authors. These commands 
%% can be used either before or after the list of corresponding authors. The
%% argument for \collaboration is the collaboration identifier. Authors are
%% encouraged to surround collaboration identifiers with ()s. The 
%% \nocollaboration command takes no argument and exists to indicate that
%% the nearby authors are not part of surrounding collaborations.

%% Mark off the abstract in the ``abstract'' environment. 
\begin{abstract}

The continuous discovery of exoplanets, each with distinctive stellar and planetary properties, along with the development of higher resolution ground and space telescopes has positioned climate evolution as a fundamental component of the study of habitability. In particular, the planet Gl 514 b, located within the habitable zone of an M0.5 dwarf star 7.62 pc from Earth, is a candidate for direct observations with future ground and space-based telescopes and therefore worthy of climate modeling. One notable aspect of this planet is its eccentricity of $e = 0.45^{+0.15}_{-0.14}$, which could affect the seasonal climate by inducing large swings in instellation over the course of an orbit. Hence, we simulate a plausible range of climates on this planet to assess the likelihood that its surface is habitable as well as estimate the surface ice coverage, which could affect the photometric signal. To perform these simulations, we use an energy balance model to explore the parameter space permitted by the observations and the allowed ranges of the obliquity, eccentricity, atmospheric \ce{CO2}, precession angle, land fraction, and land distribution. We find the planet is most likely to be in either a snowball or ice free state, but about 1.27\% of our simulations contain polar ice caps or an ice belt. A partial pressure of \ce{CO2} in the range of 7.25--9.5 bar permits the planet's surface to be habitable. These results constrain the orbital, rotational, and physical conditions required for surface habitability of Gl 514 b and will help guide future direct-imaging surveys of this and similar planets.

\end{abstract}

%% Keywords should appear after the \end{abstract} command. 
%% The AAS Journals now uses Unified Astronomy Thesaurus concepts:
%% https://astrothesaurus.org
%% You will be asked to selected these concepts during the submission process
%% but this old "keyword" functionality is maintained in case authors want
%% to include these concepts in their preprints.
%%\keywords{Classical Novae (251) --- Ultraviolet astronomy(1736) --- History of astronomy(1868) --- Interdisciplinary astronomy(804)}

%% From the front matter, we move on to the body of the paper.
%% Sections are demarcated by \section and \subsection, respectively.
%% Observe the use of the LaTeX \label
%% command after the \subsection to give a symbolic KEY to the
%% subsection for cross-referencing in a \ref command.
%% You can use LaTeX's \ref and \label commands to keep track of
%% cross-references to sections, equations, tables, and figures.
%% That way, if you change the order of any elements, LaTeX will
%% automatically renumber them.
%%
%% We recommend that authors also use the natbib \citep
%% and \citet commands to identify citations.  The citations are
%% tied to the reference list via symbolic KEYs. The KEY corresponds
%% to the KEY in the \bibitem in the reference list below. 

\section{Introduction} \label{sec:intro}

\begin{table}[!htb]
    \caption{Measured Stellar and Planetary Properties of the Gl 514 System\label{tab:properties}} 
    \begin{center}
    \begin{tabular}{@{}l l@{}}
        \hline \hline
        \textbf{Parameter} & \textbf{Value} \\  
        \hline
        \multicolumn{2}{l}{\textit{Stellar Properties}} \\
        \hline
        Mass (M$_{\star}$) & $0.510\pm 0.051 \, $M$_{\odot}$ \\ 
        Radius (R$_{\star}$) & $0.500\pm 0.047 \, $R$_{\odot}$ \\
        Luminosity (L$_{\star}$) & $0.043\pm 0.009 \, $L$_{\odot}$ \\
        \hline
        \multicolumn{2}{l}{\textit{Planetary Properties}} \\
        \hline
        eccentricity (e)  & $0.45^{+0.15}_{-0.14}$           \\
        semi-major axis (a)           & $0.422^{+0.014}_{-0.015}$ AU  \\
        Period (P)           & $140.43 \pm 0.41$ days           \\
        m\textsubscript{b}\,sin\,i\textsubscript{b} & $5.2\pm \, 0.9$ M$_{\oplus}$        \\ 
        \hline
    \end{tabular}
    \end{center}
\end{table}

The recent discovery of Gl 514 b, a nearby habitable zone (HZ) super-Earth (minimum mass $m_{b}\,\sin\,i_{b}=5.2\pm 0.9 M_{\oplus}$) exoplanet on a significantly eccentric orbit of $e = 0.45^{+0.15}_{-0.14}$ \citep{Damasso} provides an excellent benchmark case for exploring the role of high eccentricity on the climates of directly imageable habitable planets.  The planet's host star is an  M$0.5-1.0$ dwarf star at a distance of 7.62 pc, and the planet's orbital period is 140 days (semi-major axis $a=0.422\pm0.015$ AU). There is no evidence of a planetary transit in the TESS light curve \citep{Damasso_2022a}, so the radius of the planet is currently unknown. Based on these limited data, we simulated the latitudinal variations in surface temperature and ice coverage over a range of  orbital (e.g., eccentricity),
rotational (e.g., obliquity), atmospheric (e.g., partial pressure of CO$_2$), and surface
(e.g., land/water fraction) properties to quantify how they affect the climates of habitable planets. Table \ref{tab:properties} summarizes the measured parameter values for the Gl 514 b system as reported in \citet{Damasso}. 

Gl 514 b's proximity makes it a good candidate for the next generation of ground-based segmented-mirror telescopes \citep{crossfield_2013} and a potential candidate for the Habitable Worlds Observatory \citep{tuchow2024hpic}. Although the HZs of M dwarfs are generally considered poor locations for direct imaging targets, Gl 514 b lies at the outer edge of a relatively wide HZ and its large eccentricity means the angular separation at apoastron of $\sim$ 80 milliseconds of arc \citep{Damasso} is larger than the expected inner working angle of the Extremely Large Telescope \citep[ELT;  ][]{Kasper_2021}. Moreover, the planet's absolute magnitude will likely be smaller than the typical potentially habitable planet in the solar neighborhood because the planet presumably possesses a large radius and its position in the HZ could create a significantly (but hopefully not globally!) ice-covered world with a large albedo. Should the planet-to-star I-band flux ratio be greater than $\sim2\cdot 10^{-9}$, it should be detectable with the ELT \citep{Damasso}. The combination of all these factors motivates climatological studies that can help interpret future data of this potentially habitable exoplanet.

\begin{figure*}[t]
    \centering
    \includegraphics[width=\linewidth]{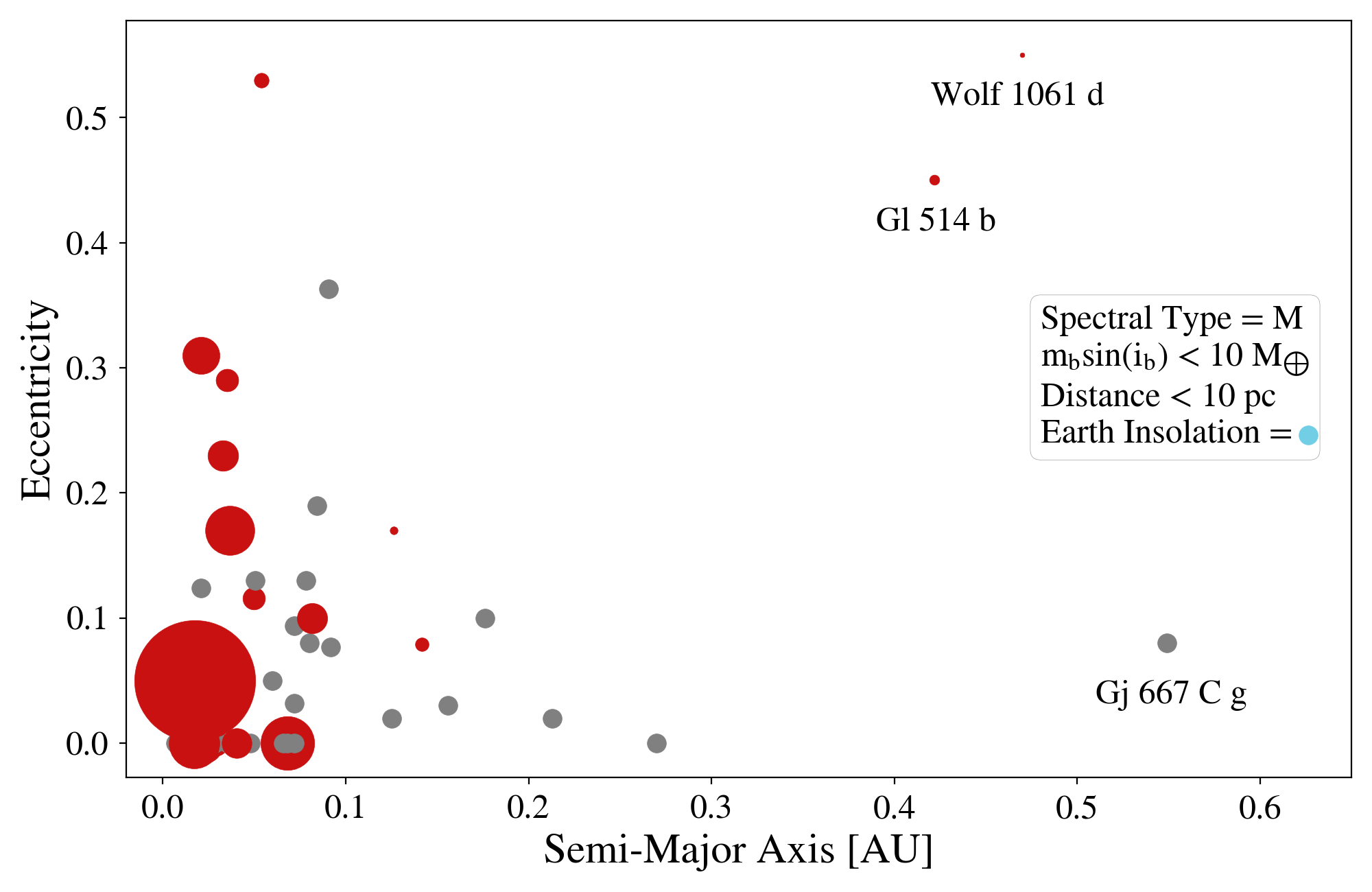}
    \caption{Eccentricity versus semi-major axis of Earth-sized and super-Earth exoplanets orbiting M type stars at distances of 10 pc or less. The size of the dot corresponds to the amount of insolation the planet receives (Earth's insolation is shown for reference). Gray  dots are exoplanets with unreported insolation due to multiple host stars. (From NASA Exoplanet Archive). }
    \label{fig:ecc}
\end{figure*}

Gl 514 b is of particular interest because its orbital properties are also relatively rare among the population of possibly terrestrial planets in the solar neighborhood. Figure \ref{fig:ecc} shows the eccentricities and semi-major axes for planets with minimum masses below 10 M$_{\oplus}$ orbiting M stars within 10 pc from the Sun \citep{ps}.  Most exoplanets with high eccentricity orbit closer to their host stars, meaning they both orbit interior to the HZ and are more difficult to study with direct imaging. The planet Wolf 1061 d (minimum mass $5.21\pm0.68$ M$_{\oplus}$) is similar to Gl 514 b, but it receives significantly less instellation than Gl 514 b, which increases its likelihood to be in a ``snowball'' state with global ice coverage.  Thus, Gl 514 b could provide the best observational constraints on the role of high eccentricity on terrestrial planet climate for the foreseeable future.

%\footnote{Data taken from \href{https://exoplanetarchive.ipac.caltech.edu/overview/trappist-1}{NASA Exoplanet Archive.}} 

With so little known about Gl 514 b today, we explored its climate with as few assumptions as possible with modern models. A planet's orbital and rotational properties, such as orbital distance, eccentricity, and obliquity, can strongly affect its climate \citep{Spiegel_2010,Armstrong_2014,Forgan_2016,Deitrick_2018,Wilhelm_2022}, so we consider the plausible range of their properties. Planet formation can generate the full range of obliquities  \citep[e.g.,][]{DONES199367,CHAMBERS2001205,Miguel_2010}, so we test all values. \cite{Palubski_2020} demonstrated that eccentric planets around M-dwarf stars can have a large segment of their orbit with temporarily habitable conditions, increasing the likelihood for maintaining water, if present, in its liquid state. Numerous studies have also shown that the distribution of land on the surface significantly impacts climate \citep[e.g.,][]{Rushby_2019}. The climate system is also driven by the abundances of radiatively active gases in the atmosphere and the distribution of clouds.  Additionally, the variation of these properties, such as Milankovitch cycles \citep{milankovic1941kanon,huybers,Wilhelm_2022} and the carbon cycle \citep{Walker1981,Kadoya_2015,Haqq-Misra2016,Abbot2016}, can also dramatically affect climate. 

A proper experiment would consider all the external and internal forcings with a fully coupled interior-climate-stellar-orbital model with high spatial resolution and direct radiative transfer, i.e., a large survey of million-year simulations with a 3D global circulation model (GCM). Such a campaign is currently computationally expensive, so we use an approximate 1D latitudinal energy balance model (EBM) because it is orders of magnitude faster. Even with this approximation, the diversity of phenomena is large and underexplored in the literature, so we limit the investigation to just the equilibrium climate of Gl 514 b over the range of orbital, rotational, surface, and CO$_2$ values the planet is likely to possess if it is terrestrial. 

EBMs can provide some insight into the climates of potentially habitable worlds, but are not ideal for predicting observables of individual planets. The model's coarse spatial resolution and simplifying assumptions regarding climate dynamics mean they are best used to identify trends and categories. GCMs are the preferred tool for observational predictions, but they are more computationally expensive by many orders of magnitude. Thus, the results of our EBM parameter sweep could help streamline GCM investigations aimed at generating robust spectral and photometric predictions for Gl 514 b. 

Prior studies of the climates of Earth-like and super-Earth exoplanets have been carried out, demonstrating the possible habitability of these systems. For example, \cite{Vladilo_2013} studied the effects of surface pressure on an Earth-like exoplanet around a Sun-like star and discussed how their results can be extrapolated to other stellar types, including M dwarfs. Similarly, \cite{Bahraminas2020} showed that surface pressure, ocean fraction, and obliquity can impact the global average temperature of planets around M-type stars. The land fraction of a planet also affects its surface habitability, where planets with high land fractions are relatively colder than those with lower land fractions \citep{Rushby_2019}. In relation to the planet's surface, the ice-albedo feedback for an exoplanet, which varies with the spectral type of the host star, maintains a warmer surface temperature for planets orbiting M dwarfs \citep{Shields_2013}. Previous studies have also shown the importance of \ce{CO2} cycling for exoplanets to maintain a stable climate with liquid water \citep{Kadoya_2014, Kadoya_2015, Haqq-Misra2016, Kadoya_2019}. 

A significant amount of research has explored the role of large eccentricity ($\ge 0.5$) on planetary habitability. \cite{Spiegel_2010} showed that a highly eccentric (up to $e = 0.83$), snowball planet can deglaciate more easily than a less eccentric planet. They also showed that the partial pressure of \ce{CO2} plays a role in increasing the surface temperature of the planet. Furthermore, the presence of a giant companion planet can perturb the eccentricity and may prevent a persistent snowball state. A companion paper by \cite{Dressing_2010} demonstrated that the allowed semi-major axis for habitable planets around a Sun-like star can be extended with larger eccentricity (studied up to 0.9) along with other orbital, rotational, and surface properties. As for Gl 514 b, \citet{Biasiotti_2024} considered eccentricities up to 0.65 and found that the planet could maintain a temperate climate for various \ce{CO2}-dominated atmospheres and ocean fractions. \citet{Biasiotti_2024}, however, did not consider the full range of rotational and surface properties, leaving lingering questions regarding the range of climates for this planet.

In this study we assess the stability of the climate and estimate the ice coverage on the surface of Gl 514 b as a function of the orientation of the rotational axis, land fraction/distribution, and atmospheric \ce{CO2} for eccentricities in the range [0.0, 0.9] (3$\sigma$). In Section \ref{sec:methods} we discuss the tools and methods used to study the surface habitability of Gl 514 b. In Section \ref{sec:results} we show the results and lastly in Section \ref{sec:discussion} we discuss the implications and significance of our results, especially the probability of stable climates with ice sheets, and discuss future work that could refine our understanding of the habitability of this exoplanet. 

\section{Methods} \label{sec:methods}

\subsection{Energy Balance Model}
\label{subsec:poise}

To understand how the known orbital parameters and unconstrained properties affect the climate of Gl 514 b, we use VPLanet \citep{Barnes_2020}, an open-source software package that can simulate the effects of eccentricity, obliquity, precession angle --- the rotational phase of a planet’s spin axis around the orbital plane, specifying where the spin axis lies relative to the orbit’s reference direction at a given time --- atmospheric \ce{CO2} concentration, and land distributions on the climate of Gl 514 b. VPLanet contains 13 physical models (``modules'') that simulate various processes related to planetary evolution (e.g., atmospheric escape, N-body evolution); however, in this project we use only VPLanet's EBM module POISE to simulate the possible climates of Gl 514 b. This EBM is based on North \& Coakley \citeyearpar{NC79}, referred to as NC79, and is a one-dimensional, latitudinal model that solves for temperature and albedo along the seasonal cycle. Another component of POISE is the simulation of dynamic ice sheets that depress the lithosphere, as described by Huybers \& Tziperman \citeyearpar{huybers}. POISE has also been compared to other EBMs via the FILLET model intercomparison project and shown to produce similar results \citep{Deitrick_2023}. For a detailed description of the POISE module, we refer the reader to \citet{Deitrick_2018} and \citet{Barnes_2020}; what follows is a brief overview and a description of new features that were not implemented in those two papers.

A key modification to the previous POISE model is the addition of multiple outgoing longwave radiation (OLR) models. The original NC79 model used a linear temperature dependence that accounts for seasonal changes in cloud cover. We have now added the \citet{Spiegel09} model that includes an infrared cooling function for a given atmospheric optical thickness. The second new model is from \citet{WK97}, hereafter referred to as WK97, which calculates the OLR as a function of the partial pressure of \ce{CO2} (p\ce{CO2}), the solar zenith angle, the surface temperature, and the surface albedo. It also calculates the top-of-atmosphere albedo as a function of the distance photons travel through the atmosphere, which is set by the solar zenith angle, the surface albedo, and the amount of \ce{CO2} and \ce{H2O} present in the atmosphere. The third new model is from \cite{Haqq-Misra2016}, which uses fourth-order polynomial fits to calculate OLR as a function of p\ce{CO2} and surface temperature. Finally, we considered the \citet{Kadoya_2019} model which also uses a simplified polynomial parameterization to calculate OLR as a function of surface temperature and p\ce{CO2}.
%, but found that the OLR as a function of surface temperature is not realistic for pressures over 5 bar, so we therefore do not consider it here.}

\begin{figure*}[t]
    \centering
    \includegraphics[width=\linewidth]{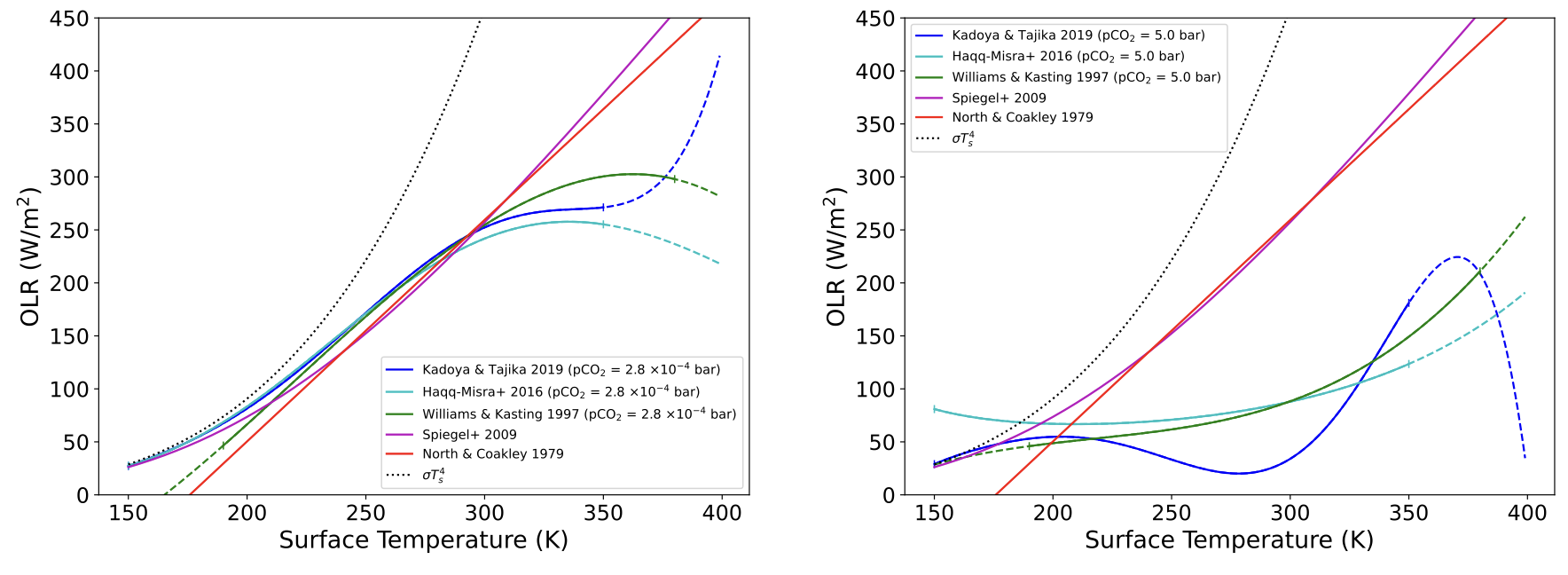}
    \caption{Different outgoing longwave radiation (OLR) models as a function of temperature. The dotted black line represents the radiation emitted by a blackbody. The solid lines represent the theoretical values for each OLR model while the dotted colored lines represent values outside of the surface temperature boundaries for each model. {\it Left:} The OLR for p\ce{CO2} = 280 ppm. {\it Right:} The OLR for p\ce{CO2} = 5 bar.}
    \label{fig:olr}
\end{figure*}

\begin{table*}[t]
    \caption{Constant Parameters\label{tab:poise}} 
    \begin{tabular}{l c l}
        \hline \hline
        \textbf{Parameter} & \textbf{Value} & \textbf{Description} \\  
        \hline
        EBM Model      &   seasonal  & Type of climate model (annual or seasonal) \\
        OLR Model      &   Williams \& Kasting 1997 & Greenhouse forcing of \ce{CO2}                     \\
        $P$\textsubscript{rot}              & 23.93 h             & Rotational Period \\
        %p$\ce{CO2}$    &   7.0 bar   & Atmospheric partial pressure of \ce{CO2}            \\
        $T$\textsubscript{glob}     &   $13.75$$^{\circ}$C & Average initial global temperature        \\
        $r$\textsubscript{snow}     &   $2.9\times10^{-5}$\, kg m\textsuperscript{-2} s\textsuperscript{-1} & Deposition rate of ice/snow to form ice sheets                                                        \\
        %C\textsubscript{land}     &   $5.00\times10^{7}$ J/K & Heat capacity of the land              \\
        %C\textsubscript{water}    &   $3.0800702\times10^7$ J/K    & Heat capacity of water                 \\
        $D$\textsubscript{mix}      & 70 m & Ocean mixing depth \\
        %D              &   0.625         &  Diffusion coefficient \\
        Ice Latitude   &   $90^{\circ}$ N/S       & Initial location of ice sheets (i.e., no ice sheets)         \\
        $t_{stop}$     &   10 Myr                   & Integration Stop Time             \\
        Output Time     &   10 kyr                   & Output interval to record on VPLanet's log file \\
        $\eta$     &   0.001                   & Scales the timestep as a fraction of the system’s \\
        & & shortest instantaneous timescale             \\
        %a\textsubscript{water}   &   0.342             & Water Albedo                                               \\
        %a\textsubscript{land}    &   0.4255            & Land Albedo                                               \\
        \hline
    \end{tabular}
\end{table*}

Figure \ref{fig:olr} shows the OLR of these models as a function of temperature, with the OLR from a blackbody included for comparison. The curves represent the theoretical values for Earth's atmosphere at 280 ppm p\ce{CO2} (left panel), and 5 bar (right panel).
%, while the overlayed data points are the results from POISE using the properties described in Table \ref{tab:poise}. 
Note that the Spiegel et al.~and NC79 models do not include the partial pressure of \ce{CO2} to calculate the OLR and hence are not suitable for our study. The WK97, \citet{Haqq-Misra2016}, and \citet{Kadoya_2019} models are very similar over a wide range of surface temperature at Earth-like p\ce{CO2} levels, but at pressures over 5 bar, the \citet{Haqq-Misra2016} model has a higher OLR than the blackbody function at very low surface temperatures, while the \citet{Kadoya_2019} model behaves unrealistically.
%and we found they generated very similar results [CHECK!]. 
We ultimately selected the  WK97 model because its simplicity reduces computational time with minimal loss of accuracy and because it behaves realistically at higher p\ce{CO2} values. It is important to note that this model's limiting boundaries for the surface temperature come from the top-of-atmosphere second-order polynomial \citep[see Appendix B of][]{WK97}. Therefore we limit the temperature to $-83.15$ ${^\circ}$C $< T < 86.85$ ${^\circ}$C. If the surface temperature of one latitude bin of the planet falls below or above these limits, we assume the planet to be in a snowball or runaway greenhouse state, respectively.

\subsection{Re-calibrated Albedo, Diffusion Coefficient, and Heat Capacities}

The original version of VPLanet \citep{Deitrick_2018} calibrated the free parameters --- the land/water/ice albedo, diffusion coefficient, and land/water heat capacities --- by reproducing the pre-industrial Earth's global surface temperature with a uniform land/ocean distribution of 25\%/75\%. This approximation was reasonable, but in this study we now also consider modern Earth's geography, as well as other land distributions. We therefore recalibrated the free parameters using the WK97 model with modern geography to increase model accuracy for a pre-industrial Earth. 

We used the land and ocean distribution data from ROCKE-3D version 2.0  with 1$^{\circ}\times$1.25$^{\circ}$ resolution \citep{Tsigaridis_2025} for the modern Earth's geography. The total land fraction of this template is 28.4\%. We collapsed the 2D data into a purely latitudinal distribution with equal surface area bins. To find the parameter combination that best reproduces modern Earth, we used MaxLEV \footnote{\url{https://rorybarnes.github.io/MaxLEV/}}, a maximum likelihood estimation tool that uses various methods; in this case we used a differential evolution optimizer \citep{Storn1997}. MaxLEV searches the six-parameter space (land/water/ice albedo, land/water heat capacities, diffusion coefficient) and minimizes a Gaussian likelihood built from a set of observed climate quantities with their uncertainties.

We first attempted to constrain the parameters against the global mean surface temperatures reported by the ROCKE-3D simulations of \citet{He2022}, which span a range of obliquities at a one-day rotation period. However, MaxLEV could not find a solution within the expected global mean temperature and OLR for modern Earth. It is worth noting that the value used here for the global mean surface temperature ($T$\textsubscript{glob}$=$13.75$^{\circ}$C) at Earth's obliquity of 23.4$^{\circ}$ does not fit on \citet{He2022}'s trend of temperature as a function of obliquity. We therefore calibrated against Earth at its present obliquity using six observables: the global mean surface temperature $T$\textsubscript{glob}$=$13.75$\pm0.1^{\circ}$C \citep{Hawkins2017}, the OLR $F$\textsubscript{OLR}$=$239$\pm$3 W/m$^2$ \citep{IPCC2021}, and the land and sea ice line latitudes at the north and south poles. For the northern land and sea ice lines we adopt 72$\pm1.72^{\circ}$N, following the 1850--1900 reconstruction of \citet{Walsh2017} where both the Greenland ice sheet boundary and the 15\% sea-ice concentration edge are near this latitude. As for the southern lines, we adopt 70$\pm1.72^{\circ}$S for the land ice line set by the Antarctic coastline, and 63$\pm1.72^{\circ}$S for the sea ice line based on the whaling records of the Antarctic sea-ice edge and historical data \citep{delaMare2009, Edinburgh2016, scar}. None of \citet{IPCC2021}, \citet{Walsh2017}, \citet{delaMare2009}, \citet{Edinburgh2016}, or \citet{scar} include uncertainties, so we set reasonable but arbitrary values since maximum likelihood calculations require uncertainties.

\begin{table*}[t]
\centering
\caption{Re-calibrated Constant Parameters\label{tab:recalib}}
\begin{tabular}{lcc}
\hline
\textbf{Parameter} & \textbf{Previous Values} & \textbf{New Values} \\
                   & \textbf{(from \citealt{Deitrick_2018})} &          \\
\hline
\multicolumn{3}{l}{\textit{Model Parameters}} \\
Ice albedo ($a_{\text{ice}}$) & 0.6 & 0.702 \\
Water albedo ($a_{\text{water}}$) & 0.263 & 0.290 \\
Land albedo ($a_{\text{land}}$) & 0.363 & 0.528 \\
Heat capacity of water ($C_{\text{W}}$) [J m$^{-2}$ K$^{-1}$] & $4.428 \times 10^{6}$ & 8.67 $\times 10^{8}$ \\
Heat capacity of land ($C_{\text{L}}$) [J m$^{-2}$ K$^{-1}$] & $1.55 \times 10^{7}$ & 2.71 $\times 10^{6}$ \\
Diffusion ($D$) & 0.58 & 0.590 \\
OLR Model       & NC79 & WK97 \\
p\ce{CO2} [ppm] & - & 280 \\
Land distribution & Uniform & Modern Earth \\
\quad Land fraction & 25\% & 28.4\% \\
\quad Ocean fraction & 75\% & 71.6\% \\
\hline
\multicolumn{3}{l}{\textit{Model Results}} \\
Global average surface temperature (T$_{\text{glob}}$) [$^{\circ}$C] & 14.7 & 13.90 \\
Outgoing longwave radiation ($F_{\text{OLR}}$) [W/m$^{2}$] & 234.04 & 233.6 \\
Northern land ice line [$^{\circ}$N] & - & 60.85 \\
Northern sea ice line [$^{\circ}$N]  & - & 47.73 \\
Southern land ice line [$^{\circ}$S] & - & 64.16 \\
Southern sea ice line [$^{\circ}$S]  & - & 64.16 \\
\hline
\hline
%\multicolumn{3}{l}{}
\textbf{Standard Values} & \textbf{Value} & \textbf{Reference} \\
\hline
Global average surface temperature (T$_{\text{glob}}$) [$^{\circ}$C] & 13.75$\pm$0.1 & \citet{Hawkins2017} \\
Outgoing longwave radiation ($F_{\text{OLR}}$) [W/m$^{2}$] & 239$\pm$3 & \citet{IPCC2021} \\
Northern land ice line [$^{\circ}$N] & 72$\pm$1.72 & \citet{Walsh2017} \\
Northern sea ice line [$^{\circ}$N]  & 72$\pm$1.72 & \citet{Walsh2017} \\
Southern land ice line [$^{\circ}$S] & 70$\pm$1.72 & \citet{scar} \\
Southern sea ice line [$^{\circ}$S]  & 63$\pm$1.72 & \citet{Edinburgh2016} \\
\hline
\end{tabular}
\end{table*}

This methodology provides a robust calibration of the VPLanet EBM that approximately reproduces the expected pre-industrial temperature (13.75$\pm$0.1$^{\circ}$C), OLR (239 W/m$^2$), and latitudinal ice line values with Earth's modern geographical land distribution. This calibration to modern Earth geography increases the confidence that our model's application to exoplanetary climate is as accurate as possible for an EBM. Table \ref{tab:recalib} shows values for the previous (using NC79, uniform land distribution) and new (using WK97, modern land distribution) calibrations.

We note that because Gl 514 is an early-type M dwarf, its stellar spectrum peaks closer to a K dwarf than a cool, late-type M dwarf. While the near-infrared output of mid-to-late M dwarfs is known to decrease surface ice albedo and weaken the runaway ice-albedo feedback \citep{Shields_2013}, these effects become critical for stars of spectral type $\sim$M3 and later. \citet{Wilhelm_2022} demonstrated that for planets orbiting K dwarfs, the value of the ice albedo does not strongly affect the climate state. Therefore, we adopt the ice albedo from our calibration rather than an isolated modification of this parameter.

We find that the recalibration is not a perfect match to the pre-industrial Earth, with the northern sea ice line a particularly poor fit. We attribute this discrepancy to the model's simplistic representation of the Earth's actual complex and non-linear climate system. Nonetheless, we prefer this approach to others that employ a uniform land distribution as here we are calibrating to actual Earth data as opposed to a similar, but idealized, version of Earth.

\subsection{Simulations}

We use the stellar, planetary, and orbital parameters of the Gl 514 system from \cite{Damasso}. As for the unknown parameters (e.g., obliquity), we used POISE to explore different configurations to understand and constrain the possible conditions needed for the planet to be habitable. We also analyzed the ice sheet coverage of the planet, categorizing it into four possible states: snowball, polar ice cap(s), ice belt, and ice free. As a planet transitions into a snowball state, the expansion of the ice sheets triggers a runaway positive feedback loop by increasing the surface albedo, thus rapidly lowering surface temperatures. Once the planet is glaciated, it can be difficult for it to recover from such a state unless it has a mechanism (e.g., carbonate-silicate weathering cycle) to prevent further cooling \citep{Haqq-Misra2016}. As for an ice belt state, the planet must have a high obliquity where the majority of the incident stellar flux hits the poles rather than  the equator \citep{Williams_2003, Kilic_2017, Rose_2017, Kilic_2018, Colose_2019}. 

\begin{table*}[t]
    \caption{Summary of Simulations with Fixed and Varied Parameters\label{tab:simulations}} 
    \begin{tabular}{l c l}
        \hline \hline
        \textbf{Name (N simulations)} & \textbf{Description} & \textbf{Parameters} \\
        \hline
        \multirow{4}{0.2\textwidth}{Set A (N = 15,555)}      &     &  $e=0.45$\\
        & Find p\ce{CO2}, $\varepsilon$, and $\psi$ that reproduce a surface &  p\ce{CO2}=$[7.25-8.5]$ bar \\
        & temperature similar to pre-industrial Earth &  $\varepsilon=[0^{\circ}-90^{\circ}]$ \\
        & and polar ice caps &  $\psi=[0^{\circ}-180^{\circ}]$ \\
        \hline
        \multirow{4}{0.2\textwidth}{Set B (N = 201)}  &        &   $e=[0.0, 0.45, 0.9]$   \\
        & Study the latitudinal temperature  & p\ce{CO2} = 8.375 bar \\
        & variations for various eccentricities &  $\varepsilon=3.0^{\circ}$ \\
        &  &  $\psi=90^{\circ}$ \\
        \hline
        \multirow{4}{0.2\textwidth}{Set C (N = 40,804)}   &    & $e=[0.15-0.9]$  \\
        &  Parameter sweep of eccentricity,  & p\ce{CO2} = $[8.0-11.0]$ bar \\
        & p\ce{CO2}, and obliquity &  $\varepsilon=[0^{\circ}-90^{\circ}]$ \\
        &  &  $\psi=45^{\circ}$\\
        \hline
        \multirow{4}{0.2\textwidth}{Set D (N = 20,336)} &  & $e=0.45$ \\ 
        & Parameter sweep of p\ce{CO2},  & p\ce{CO2} = $[7.0-10.0]$ bar \\
        & obliquity, and land fraction &  $\varepsilon=[0^{\circ}-90^{\circ}]$ \\
        &  &  L\textsubscript{frac} = $[10\%-90\%]$ \\
        \hline
        \multirow{5}{0.2\textwidth}{Set E (N = 56,420)} & & $e=[0.0-0.9]$ \\
        &  Parameter sweep of p\ce{CO2}, &  p\ce{CO2} = $8.375$ bar \\
        & obliquity, and type of distribution  & $\varepsilon=[0^{\circ}-90^{\circ}]$ \\
        &  &  $\psi=[0^{\circ}-360^{\circ}]$, \\
        &  &  L\textsubscript{dist} = [equatorial, polar, \\
        &  &  \quad \quad \quad \quad random, modern] \\
        \hline
        \textbf{Total simulations: 133,316} & & \\
        \hline
    \end{tabular}
\end{table*}

For this study, we perform 5 different sets of simulations labeled A--E (see Table \ref{tab:simulations}). Our goal with Set A is to identify which combinations of obliquity ($\varepsilon$), precession angle ($\psi$), and p\ce{CO2} produce an average global temperature of 286.9 K (13.75$^{\circ}$C), assuming the nominal eccentricity. To simplify our simulations, in all cases we suppress the stellar torque to prevent axial precession. In other words, the climate obliquity precession parameter \citep[COPP; ][]{Deitrick_2018a}, given by 
\begin{equation}
    \chi = e\ \sin(\varepsilon)\ \sin(\varpi + \psi),
\end{equation}
remains constant, where $\varpi$ is the longitude of periastron or the angle from the reference direction (i.e., vernal equinox in POISE) to the pericenter. In our code, $\varpi$ is calculated as

\begin{equation}
    \varpi = \varpi_0 + \psi + \pi
\end{equation}
where $\varpi_0$ is the initial longitude of pericenter, which in our simulations is 0$^{\circ}$. In this set, the distribution of land and ocean is held constant at 28.4\%/71.6\% as a function of latitude.

In Set B the value of p\ce{CO2} is fixed and we survey different eccentricities to study their effect on the global temperature and ice sheets. We use a partial pressure of \ce{CO2} and rotational configuration from Set A that produced the pre-industrial Earth-like surface temperature as well as two polar caps (8.375 bar of \ce{CO2}, $\varepsilon=3.0^{\circ}$, $\psi = 90^{\circ}$). We survey values of $e = [0.0, 0.9]$ as the goal of this set is to understand the climate at values within $3\sigma$ from the mean eccentricity as constrained by \cite{Damasso}.
%, however for eccentricities of $e=0.0-0.449$, the simulations fall out of the surface temperature boundaries of the WK97 model, indicating the planet has collapsed into a snowball state. Therefore, we only show the results for eccentricities above 0.449.}

In Set C we run a parameter sweep for continuous values of $e =[0.15, 0.9]$, $\varepsilon=[0^{\circ}, 90^{\circ}]$, and p\ce{CO2}=$[8.0, 11.0]$ bar to study their effects on the global average temperature for specific values of p\ce{CO2}. We limit the range of p\ce{CO2} in this set to only those values that can result in temperatures within the bounds of the WK97 model. Depending on the configuration of the other orbital parameters, the temperature of the planet could still exceed the boundaries of the WK97 model and the planet can be inferred to be a snowball for $T<-83.15$$^{\circ}$C or a runaway greenhouse for $T>86.85$$^{\circ}$C.

In Set D we study how the land fraction of the planet affects habitability in terms of p\ce{CO2} and $\varepsilon$. As in Set C, we limit the p\ce{CO2} range to those values that permit temperatures within the bounds of the WK97 model. Note also that POISE becomes numerically unstable if a latitude possesses 100\% land or ocean, so we also limit the land and ocean fractions to be in the range 1\%--99\% per bin. We assume a constant land fraction as a function of latitude.

Lastly, Set E studies how the land distribution affects the end state and surface temperature of the planet. We consider four different configurations: all land near the equator, all land split evenly between the poles, randomly  distributed land, and Earth's modern distribution. All four modes have a global land fraction of 28.4\% for consistency. The equatorial distribution is therefore a slab of land between 16$^{\circ}$N and 16$^{\circ}$S, whereas the polar land distribution consists of two continents located on each pole, both extending 44.3$^{\circ}$ towards the equator. The random option allows variations of up to 20\% between adjacent latitudinal bins, and we used the same random land distribution for all cases.

Also note that because we use a uniform distribution of water and land for all sets except Set E, the values for $\psi$ are only probed from $[0^\circ$--$180^\circ]$ due to symmetry. A summary of the simulations can be found in Table \ref{tab:simulations}. In total, we perform over 130,000 simulations of Gl 514 b's climate.

\section{Results}
\label{sec:results}

%\subsection{Static Simulations}
%\label{subsec:static}

\begin{figure*}[t]
    \centering
    \includegraphics[width=\linewidth]{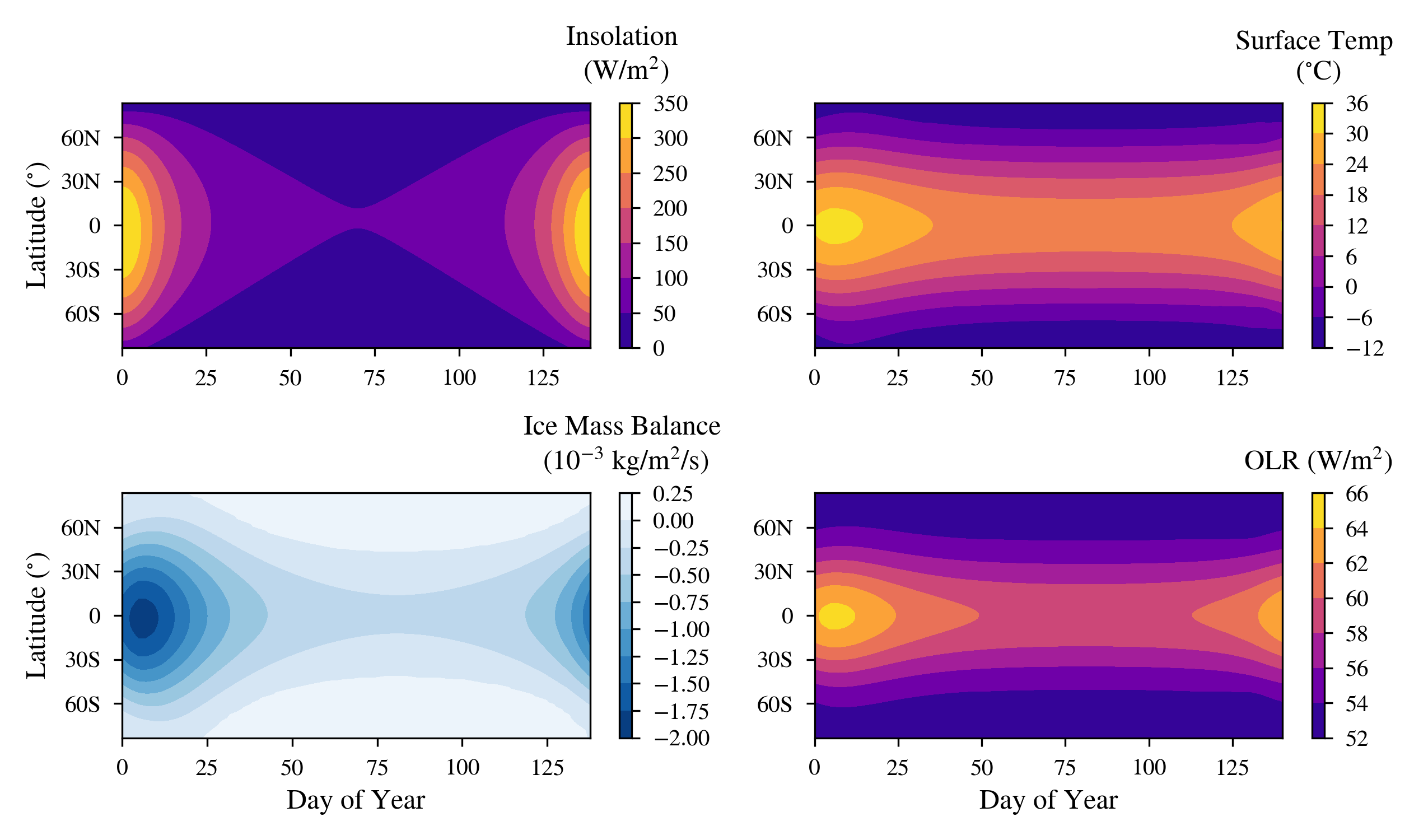}
    \caption{An individual simulation from Set A showing the annual latitudinal climate of Gl 514 b ($\varepsilon=3.0^{\circ}$, $\psi = 90^{\circ}$, and p\ce{CO2} = 8.375 bar). \textit{Top left}: Insolation received by the planet. \textit{Top right}: Surface temperature. \textit{Bottom left}: Ice mass balance. A positive value represents ice accumulation while a negative value represents ice melting. \textit{Bottom right}: Outgoing longwave radiation.} 
    \label{fig:static_climate}
\end{figure*}

We first consider the results of Set A and find that surface habitability requires p\ce{CO2} $>$ 7.5 bar. Out of 15,555 simulations, 230 finished with polar ice caps. However, the closest global average surface temperature to pre-industrial Earth was 8.0$^{\circ}$C. To showcase an example with polar caps, Figure \ref{fig:static_climate} presents the climate of the exoplanet during its last orbital period if $\varepsilon=3.0^{\circ}$, $\psi = 90^{\circ}$, and p\ce{CO2} = 8.375 bar. The mean global temperature of the planet is 8.0$^{\circ}$C. It is evident that the properties presented in each of the panels have a strong correlation, as expected. Another feature to note is the ice mass balance, which measures the accumulation of ice. The positive values represent ice being accumulated, while negative values represent ice being lost. Because the insolation at the poles is low, ice can accumulate to form ice sheets, while at lower latitudes the water is in liquid form. 

\begin{figure*}[!t]
    \centering
    \includegraphics[width=\linewidth]{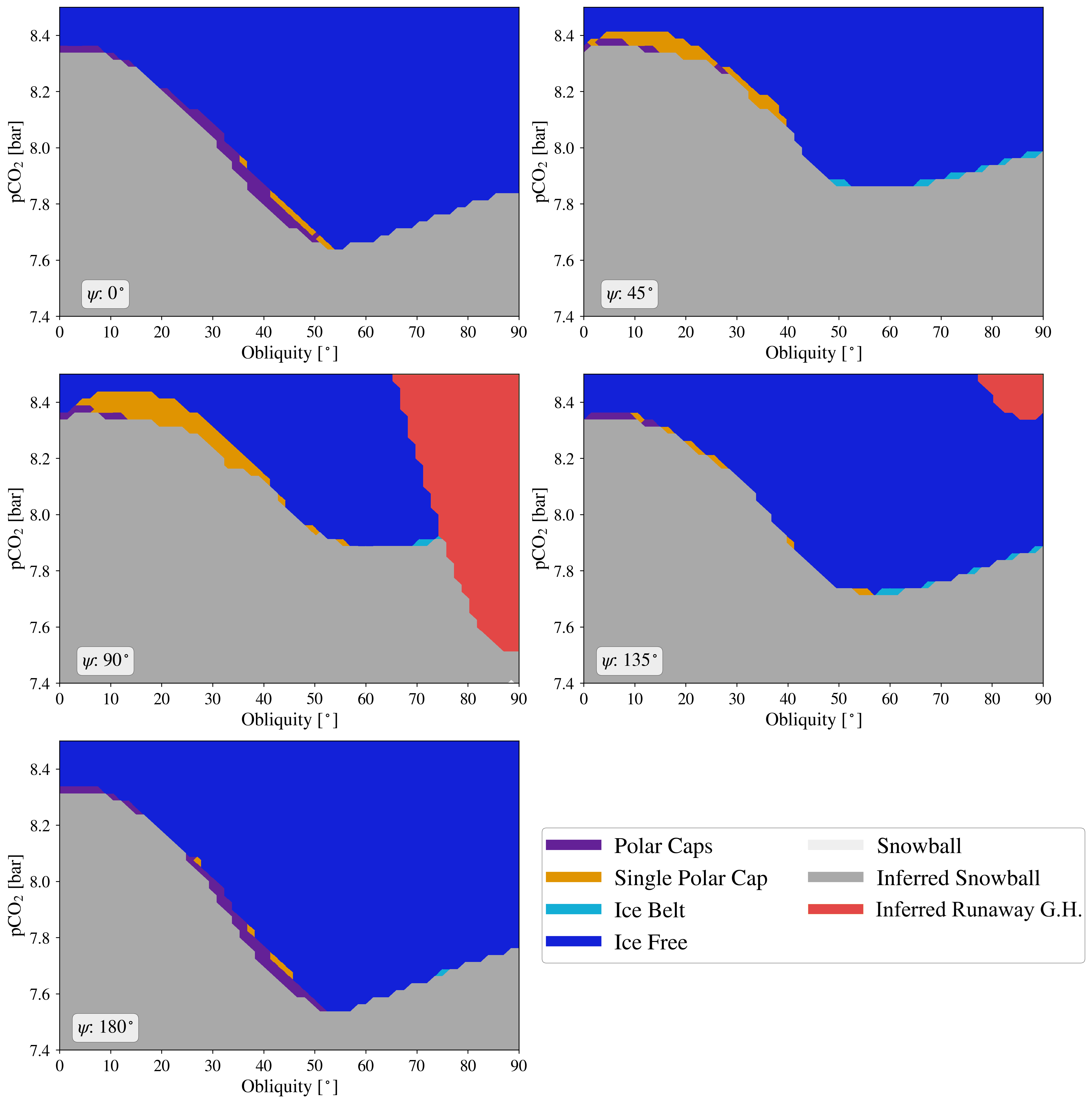}
    \caption{Final climate states for all the simulations in Set A as a function of precession angle, atmospheric \ce{CO2}, and obliquity. Note that the simulations that were out of bounds are classified as inferred snowball and inferred runaway greenhouse when exceeding the OLR model boundaries.} 
    \label{fig:setA_subplots}
\end{figure*}

Figure \ref{fig:setA_subplots} shows the final ice states for all the simulations of Set A as a function of their atmospheric \ce{CO2} pressure, obliquity, and precession angle (0$^{\circ}$, 45$^{\circ}$ in the top row; 90$^{\circ}$, 135$^{\circ}$ in the middle row, and 180$^{\circ}$ in the bottom row). The ice states are classified as polar caps, single polar cap, ice belt, ice free, snowball, inferred snowball (i.e., the temperature was below the WK97 limit), and inferred runaway greenhouse (i.e., the temperature was above the WK97 limit). 

The formation of polar caps tends to occur at low obliquities, as expected from \citet{Wilhelm_2022}, who also used POISE. Yet because Gl 514 b has a high eccentricity, the stellar flux varies by a factor of $\sim$7 between periastron and apoastron. Hence $\psi$ (and consequently $\varpi$) impacts the formation of polar caps at higher obliquities (i.e., $>40^{\circ}$). When $\psi=90^{\circ}$, the planet is experiencing the northern winter solstice at pericenter, which allows for the formation of a single polar cap in the southern hemisphere at high obliquities, as shown in the middle-left panel of Figure \ref{fig:setA_subplots}. Moreover, for higher obliquities the planet goes into an inferred runaway greenhouse state as the northern hemisphere experiences a hot winter, which does not allow the planet to cool off. At $\psi=135^{\circ}$ the northern hemisphere experiences its winter 45$^{\circ}$ after periapsis, which also does not allow the planet to cool. As for ice belts, because the stellar flux variation in the orbit impacts the poles more than the equator, they are only found at high obliquities, also as expected from \citet{Wilhelm_2022}.

Since all simulations use a uniform land distribution, the northern and southern hemispheres are identical. Consequently, we expect a $180^{\circ}$ shift in $\psi$ to yield statistically identical climate states, as this shift only swaps which hemisphere experiences a given season at periastron (e.g., comparing $\psi = 90^{\circ}$ to $\psi = 270^{\circ}$, or $\psi = 0^{\circ}$ to $\psi = 180^{\circ}$). We note, however, some small differences between these 180$^{\circ}$ symmetry pairs, especially near climate state transitions. This asymmetry arises because each simulation begins at the southern summer solstice, and so the differences result from the planet either moving towards or away from periastron. This setup can cause small but significant differences that tip the climate into one state or the other. For simulations far from the climate state transitions, the expected 180$^{\circ}$ precessional symmetry holds true across the parameter space.

\begin{figure*}[t]
    \centering
    \includegraphics[width=\linewidth]{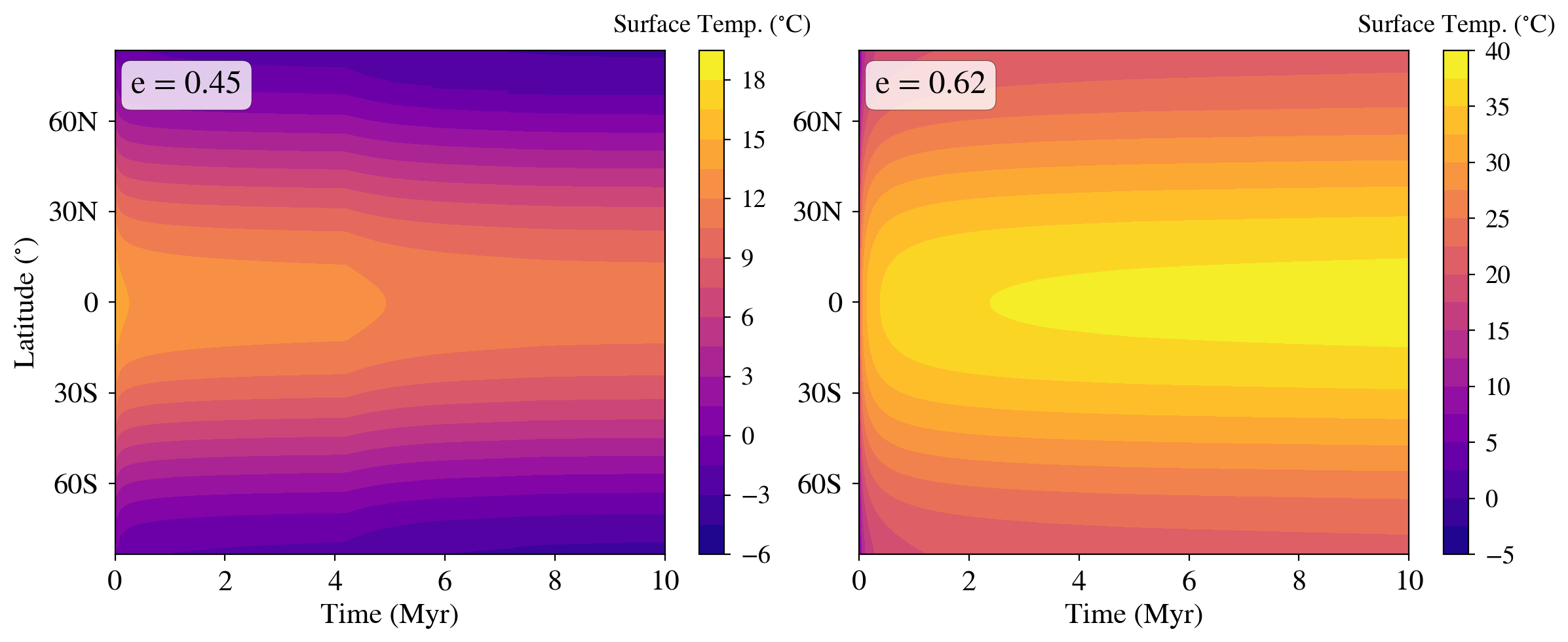}%scale=0.65
    \caption{Set B latitudinal annual temperature variation of Gl 514 b after 10 Myr evolution for two different orbital configurations. The nearly horizontal contours after $\sim$5 Myr indicate the model prediction has stabilized.
    \textit{Left panel}: Surface temperature (in $^{\circ}$C) for $e=0.45$. \textit{Right panel}: $e=0.62$. Note that the color bar for the right panel is not the same due to the large difference in surface temperature between each simulation.}  
    \label{fig:static_survey}
\end{figure*}

Turning to Set B, Figure \ref{fig:static_survey} shows the annual variation of the surface temperature over 10 Myr for two different eccentricities of 0.45 and 0.62. For eccentricities below 0.45, the temperature drops quickly below $-83.15$$^{\circ}$C, going out of the bounds of the WK97 OLR model, indicating the planet has entered the full snowball state. This outcome is due to the planet's relatively low eccentricity, which lowers the orbit-averaged instellation. We therefore performed 100 simulations to find the eccentricity at which the surface temperature is within the acceptable range of the WK97 OLR model. The lowest eccentricity allowed was $e=0.45$, which coincidentally is the same as the nominal value. These results suggest extremely low temperatures for eccentricities near 0. Conversely, at $e=0.9$ the surface temperature exceeded the WK97 upper boundary, so we performed 100 simulations to find the value where the climate evolves up to the stop time. The maximum eccentricity allowed with the WK97 model for this configuration of parameters is 0.62. The left panel shows the surface temperature for the value of $e$  with a mean global temperature of 8.0$^{\circ}$C, resulting in a polar ice cap state. The right panel shows the results for an eccentricity of 0.62, which has a mean global temperature of 34.1$^{\circ}$C and a completely ice free surface.

\begin{figure*}[t]
    \centering
    \includegraphics[width=\linewidth]{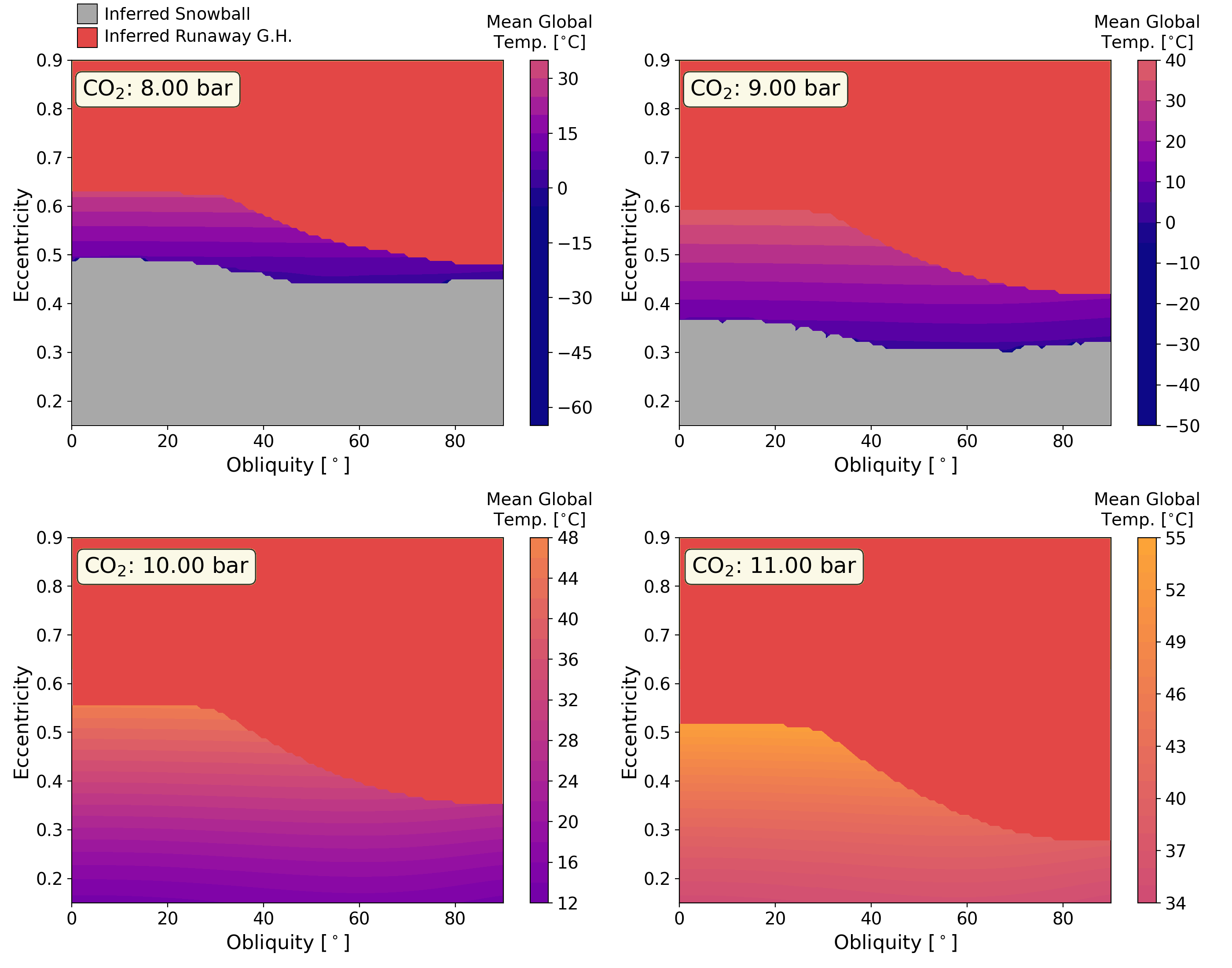}%scale=0.55
    \caption{Contour plots of Set C showing the annual mean global temperature of the planet for various $e$ and $\varepsilon$ configurations at different atmospheric \ce{CO2} levels in bar. Simulations whose temperatures fall outside the model bounds are classified as inferred snowballs (dark gray) or inferred runaway greenhouse gas (orange red).}
    \label{fig:statictglob}
\end{figure*}

In Set C we study the effect of eccentricity and obliquity on the annual mean global temperature for a range of p\ce{CO2} values. Out of 40,804 simulations, 66 finished with polar caps and 10 with ice belts, while the rest were ice free or snowballs. Figure \ref{fig:statictglob} shows four contour plots of the mean global temperature for $e=[0.15$--$0.90]$ and $\varepsilon=[0^{\circ}$--$90^{\circ}]$. The plot shows that increasing p\ce{CO2} produces temperate climates at lower eccentricities. Increasing p\ce{CO2} from 8.0 bar to 11.0 bar makes the maximum eccentricity for surface habitability drop from $e\sim0.63$ to 0.51.

\begin{figure*}[t]
    \centering
    \includegraphics[width=\linewidth]{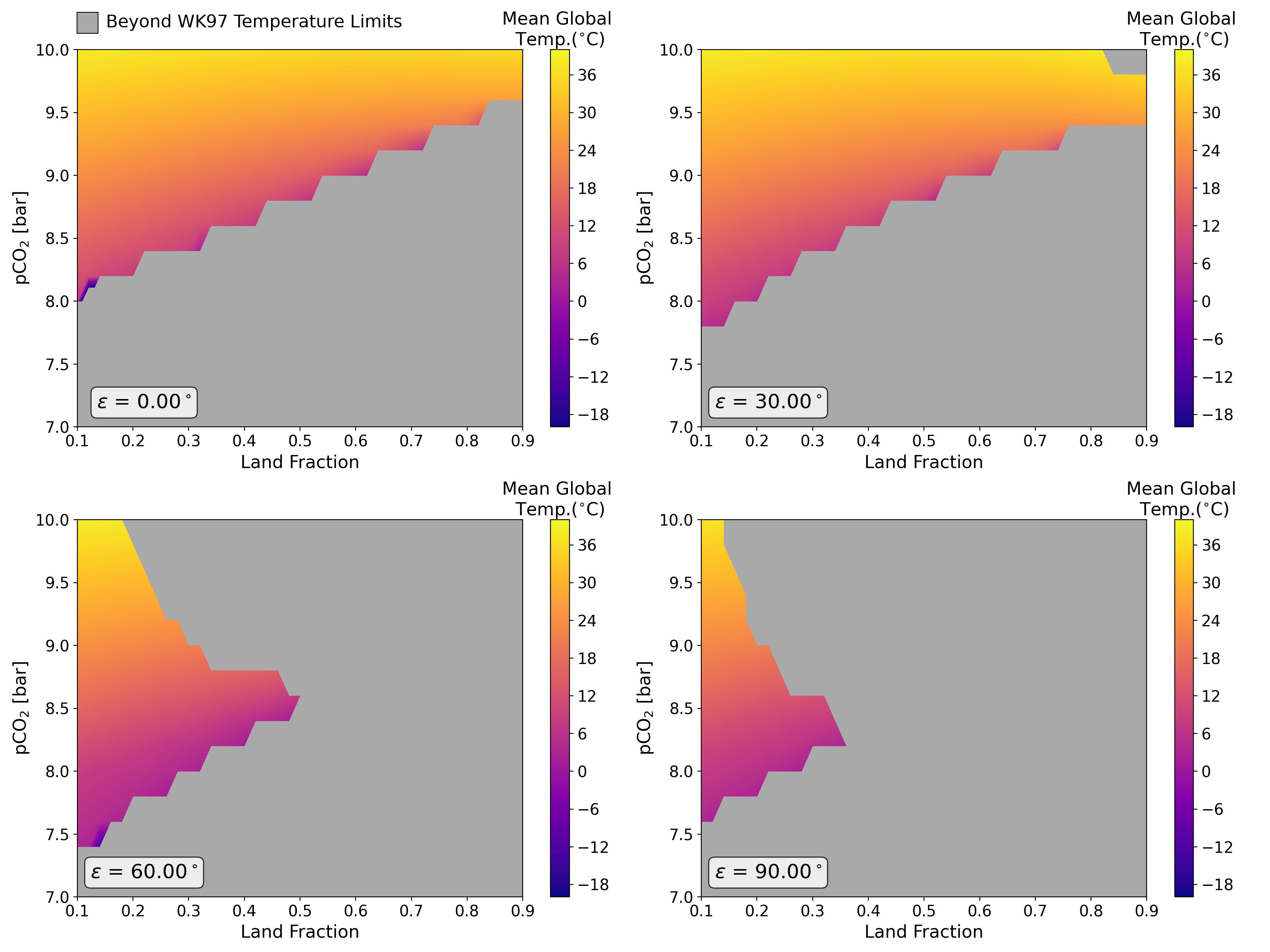}%scale=0.6
    \caption{Annual mean global surface temperature for different configurations of p\ce{CO2} and land fraction at 4 different $\varepsilon$ values for Set D. Simulations where the temperature exceeded the boundaries of the WK97 model are shown in dark gray.}
    \label{fig:landfrac}
\end{figure*}

Set D explores the effect of land fraction for different p\ce{CO2} and $\varepsilon$. In Figure \ref{fig:landfrac}, the plots show the annual mean global surface temperature for different $\varepsilon$ as a function of p\ce{CO2} and land fraction. The dark gray region shows where the surface temperature was below the bounds of the WK97 model. At higher obliquities (60$^{\circ}$ and 90$^{\circ}$), the planet maintains surface habitability at lower p\ce{CO2} values because increased polar insolation suppresses ice cap formation, reducing planetary albedo via the ice-albedo feedback. Although higher land fractions cool the surface because of their higher albedo, the direct insolation at the poles at high obliquities raises the mean surface temperature, superseding the cooling effect from higher land fractions. However, the increased temperature due to high obliquities cannot by itself warm the planet to a temperate climate if the p\ce{CO2} is low, regardless of the land fraction. The minimum p\ce{CO2} that can produce a temperate climate is 7.45 bar with $\lesssim$14\% land fraction and 60$^{\circ}$ obliquity. In contrast, low-obliquity cases ($\varepsilon$ = 0$^{\circ}$ and 30$^{\circ}$) require substantially higher \ce{CO2} to prevent a permanent snowball (dark gray regions), as permanent polar ice caps increase reflectivity. Additionally, increasing land fraction tends to decrease the temperature of the surface due to the relatively high albedo compared to water, consistent with \citet{Rushby_2019}.

%The amount of \ce{CO2} required for surface habitability appears to reach a minimum near an obliquity of $\sim 60^{\circ}$ (bottom left panel). This obliquity of 60$^{\circ}$ falls within the region of climate instability where huge seasonal swings along with equatorial heat transport produce rapidly changing climate states \citep{Ferreira_2014}. 

\begin{figure*}[t]
    \centering
    \includegraphics[width=\linewidth]{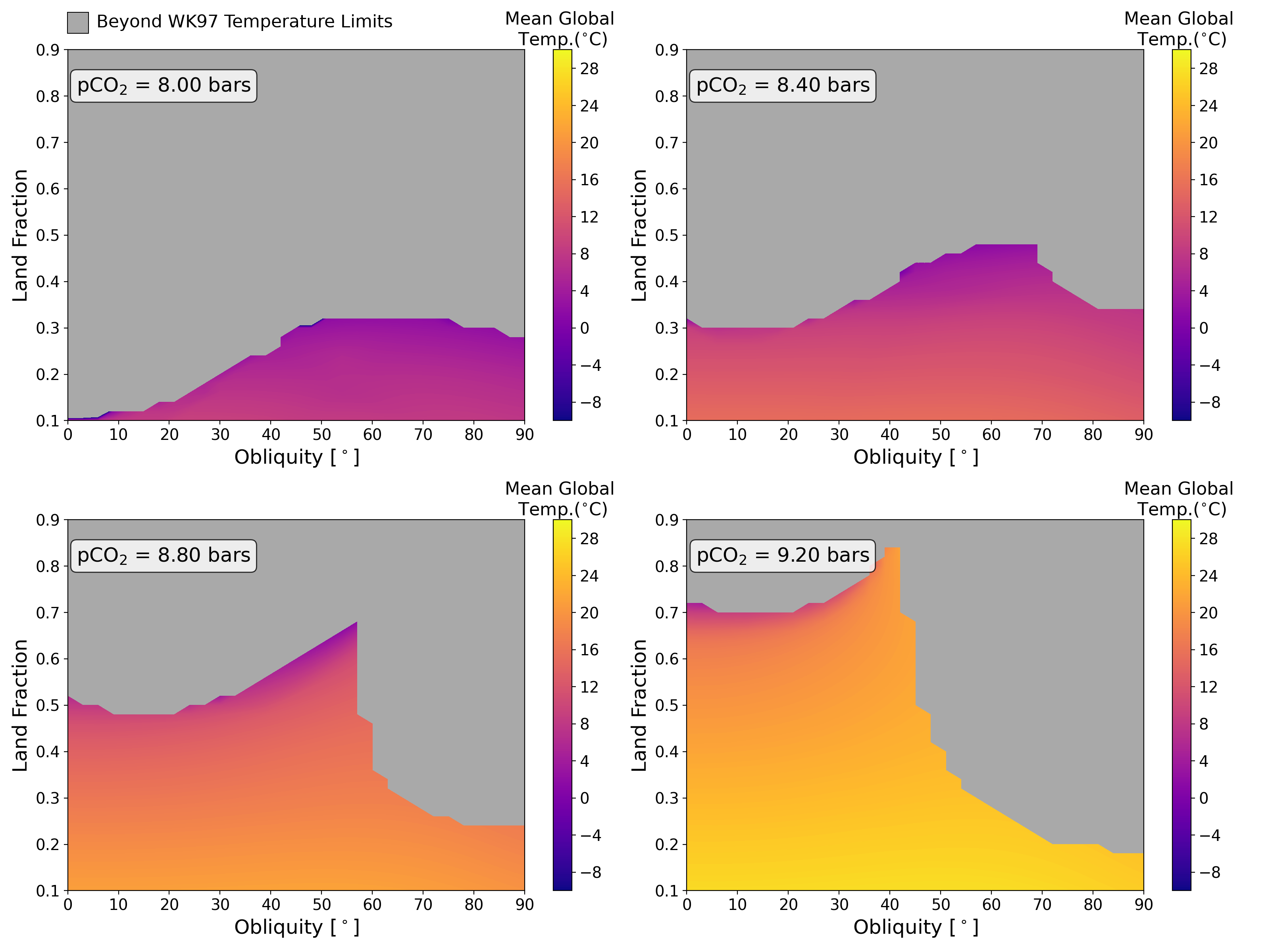}%scale=0.6
    \caption{Annual mean global surface temperature for different configurations of land fraction and obliquity at 4 different p\ce{CO2} values for Set D. Simulations where the surface temperature exceeds the WK97 boundaries are shown in dark gray.} 
    \label{fig:setD_4panel}
\end{figure*}

Figure \ref{fig:setD_4panel} shows the annual mean global surface temperature for different land fraction and obliquity values at four different levels of atmospheric \ce{CO2}. As p\ce{CO2} increases, warmer climates occur for all obliquities. At 9.20 bar, the surface temperature falls below the WK97 cold limit at low obliquities and high land fractions, while at higher obliquities it exceeds the hot limit even at low land fractions, similar to Figure \ref{fig:landfrac}. The majority of the end states for this set of simulations are either snowball or ice free, with some polar ice caps (N = 99) at $\varepsilon\approx90^{\circ}$.

\begin{figure*}[t]
    \centering
    \includegraphics[width=\linewidth]{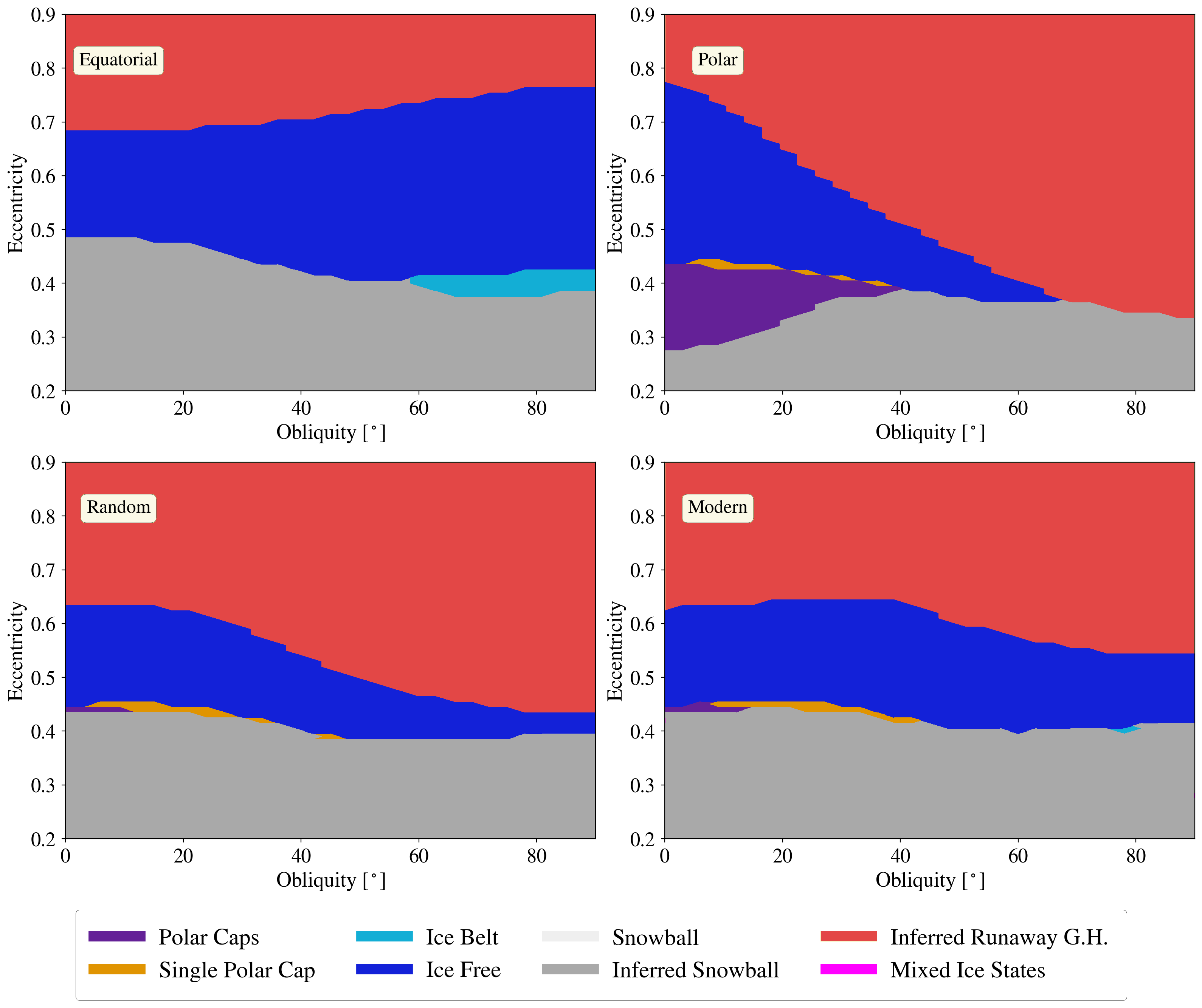}%scale=0.9
    \caption{Set E comparison of final ice coverage of a 28.4\% land fraction for equatorial (top left), polar (top right), random (bottom left), and modern (bottom right) land distributions as a function of eccentricity vs. obliquity at p\ce{CO2} = 8.375 bar and $\psi=90^{\circ}$.} 
    \label{fig:landdist}
\end{figure*}

Finally, Set E considers how the distribution of land affects the climate for different eccentricities, obliquities, and p\ce{CO2}. Figure \ref{fig:landdist} shows the ice states as a function of eccentricity vs. obliquity with p\ce{CO2} = 8.375 bar for four different land configurations: equatorial, polar, random, and modern Earth. We only plot the results for one value of p\ce{CO2} and $\psi$ as the overall trend of ice states stays fairly constant for each p\ce{CO2} and $\psi$ value.

Overall, polar caps or ice belts are possible for eccentricities below 0.48, which is close to the best-fit eccentricity. The majority of habitable cases in the entire set were ice free, with 1.4\% being polar caps, and less than 1\% ice belts. The panels of Figure~\ref{fig:landdist} show several noteworthy features. For eccentricities $> 0.78$ the temperature exceeds the upper limit of the WK97 model and the planet is likely in a runaway greenhouse, similar to our uniform land distribution simulations in Figure \ref{fig:statictglob}. When comparing the impact of the land distribution, a clear distinction in ice coverage is present between the equatorial and polar land masses, with the former able to form ice belts and the latter ice caps. This difference is not surprising since the land albedo is higher than the ocean albedo. Polar land masses may even form  ice caps for  $\psi >45^{\circ}$ in the polar distribution.

We conclude that the land distribution for a specific land fraction is not a critical factor for the surface habitability of Gl 514 b. However, it is worth noting that it could affect the photometric signal due to the differences of the surface reflectance, especially if the eccentricity is between 0.28 and 0.48. 

\begin{figure*}[t]
    \centering
    \includegraphics[width=\linewidth]{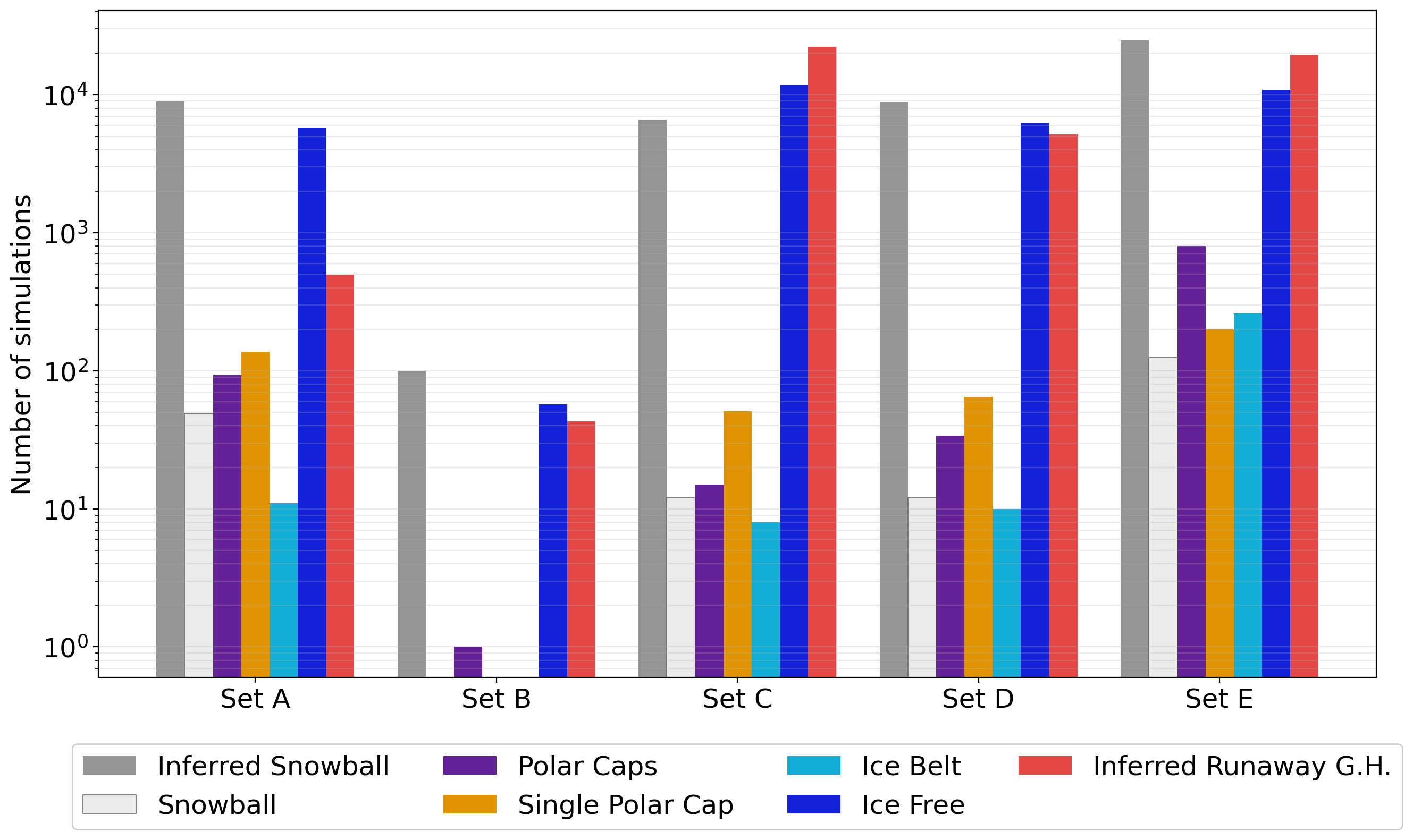}%scale=0.9
    \caption{Histogram of the various possible climate states for each Set. Note the scale is logarithmic due to the large difference between the lowest and highest values.} 
    \label{fig:histograms}
\end{figure*}

\begin{table}[!htb]
    \centering
    \caption{Climate state classifications across all sets ($N = 133{,}316$).}
    \label{tab:all_classifications}
    \begin{tabular}{lr}
        \hline
        \textbf{Climate State} & \textbf{Percentage} \\
        \hline
        Inferred Snowball     & 36.97\% \\
        Snowball              & 0.15\% \\
        Polar Caps            & 1.05\% \\
        Ice Belt              & 0.22\% \\
        Ice Free              & 25.98\% \\
        Inferred Runaway G.H. & 35.63\% \\
        \hline
    \end{tabular}
\end{table}

Figure \ref{fig:histograms} summarizes the total number of climate states per Set, while Table \ref{tab:all_classifications} shows the results in percentage for the possible climate states across all sets. About 1.27\% of the simulations resulted in single/bipolar caps or ice belts, yet the majority of the simulations were either inferred snowball, ice free, or inferred runaway greenhouse. All sets but B allow the planet to possess a single cap.

\section{Discussion and Conclusion}
\label{sec:discussion}

We used VPLanet's one-dimensional seasonal energy balance model POISE to simulate the coupled effects of orbital, rotational, atmospheric, and surface parameters on the climate of Gl 514 b. After performing more than 130,000 simulations, we found that surface habitability is possible, but it is strongly dependent on atmospheric \ce{CO2}, eccentricity, obliquity, and land fraction, while precession angle and land distribution have a lower impact. The planet is mostly found in either an ice free or snowball state, with only 1.27\% of simulations producing stable polar caps or ice belts. We caution, however, that this fraction is not intended to represent the actual probability that the planet possesses partial ice coverage; it is only the fraction in the limited parameter space we explored. Nonetheless, our results suggest the planet could be habitable, in agreement with \citet{Biasiotti_2024}. Additionally, we find that a p\ce{CO2} range of 7.25--9.5 bar is the most likely to produce a temperate climate. Such massive \ce{CO2} atmospheres are physically plausible for terrestrial planets orbiting M dwarfs like Gl 514 \citep{vonParis2013, Krissansen2023}.

%The range of p\ce{CO2} where polar caps and ice-belts were present was 7–9 bar, while eccentricities near 0.42 were most favorable for partial ice coverage. Higher \ce{CO2} concentrations can offset the cooling effect of large land fractions, and obliquities near 60$^{\circ}$ introduce a region of climatic instability. These results demonstrate that Gl 514 b can plausibly sustain pre-industrial Earth surface temperatures. Set C discussion - The majority of the simulations with polar ice caps have an eccentricity between 0.5 and 0.6.

In Set A we found orbital, rotational, and p\ce{CO2} configurations that resulted in a temperate planet, and even Earth-like polar ice caps. The main takeaway from this experiment is that the surface of Gl 514 b can be habitable. It can also have the same ice state as pre-industrial Earth with an 8.375 bar \ce{CO2} atmosphere for a range of obliquities between 0$^{\circ}$ and 40$^{\circ}$. Holding p\ce{CO2} fixed at 8.375 bar, we calculated the latitudinal variation of surface temperature for $e=[0.45, 0.62]$ in Set B where the low eccentricity simulations resulted in inferred snowballs while high eccentricity resulted in inferred runaway greenhouse states. This result highlights the climate's sensitivity to the orbital eccentricity.

In Set C we performed a parameter sweep for eccentricities and obliquities to study their effect on the mean global temperature for 8.0--11.0 bar p\ce{CO2} atmospheres. The simulations covered a wide range of ice states from snowball to ice free. Additionally, the simulations of Set D demonstrated how an increase in land fraction resulted in colder climates, but increasing p\ce{CO2} can supersede that effect, as studied and concluded by \cite{Rushby2017} and \citet{Macdonald_2022}. The trend follows a similar pattern for different obliquities, although at p\ce{CO2} values higher than $\sim$8.40 bar, lower land fractions are allowed. The last parameter sweep (Set E) revealed that land distribution does not significantly impact the climate of the planet unless the land is concentrated at the equator or poles. 

In summary, we have found that, although there are numerous properties of Gl 514 b that are different from Earth's, some parameter configurations yield temperatures that are close to the pre-industrial Earth. Thus, the planet's surface could be habitable, perhaps even ``Earth-like''. However, there are important caveats regarding these findings.

First and foremost, the 1D climate model relies on various simplifying assumptions, adopted as a trade-off for computational efficiency, that could produce spurious results. The model is cloud-free, the fraction of land can only be distributed per latitude (not longitudinally), and, although the WK97 model contains other gases (i.e., \ce{N2}, \ce{O2}, \ce{H2O}), the only atmospheric gas that can vary in POISE is \ce{CO2}. Moreover, we used Earth's properties as summarized in Table \ref{tab:poise}. As shown in Table \ref{tab:recalib}, our model parameters tended to predict lower-latitude ice lines than on Earth, suggesting our model likely predicts more ice coverage than in reality. Future studies could explore beyond these assumptions with either EBMs or GCMs.

When tuning our simulation to modern Earth's land distribution, our priority was to achieve a global mean surface temperature of 13.75$^{\circ}$C and an OLR of 239 W/m$^2$.  While we succeeded in this endeavor, we note that, compared to our original values, the heat capacity of water (C$_{W}$) increased significantly while the heat capacity of land (C$_{L}$) decreased by almost an order of magnitude (see Table \ref{tab:recalib}). Our new C$_{W}$ is almost 3 orders of magnitude greater than the C$_{L}$. In other successful EBMs, C$_{W}$ exceeds C$_{L}$ by factors of $\sim$20 \citep{Rushby_2019} to $\sim$40 \citep{Haqq-Misra_2022}. However, those models either used NC79, which does not explicitly take into account the effect of atmospheric \ce{CO2} on the OLR, or produced an OLR far greater than Earth's expected value. 

\begin{figure*}[htbp!]
    \centering
    \includegraphics[width=0.6\linewidth]{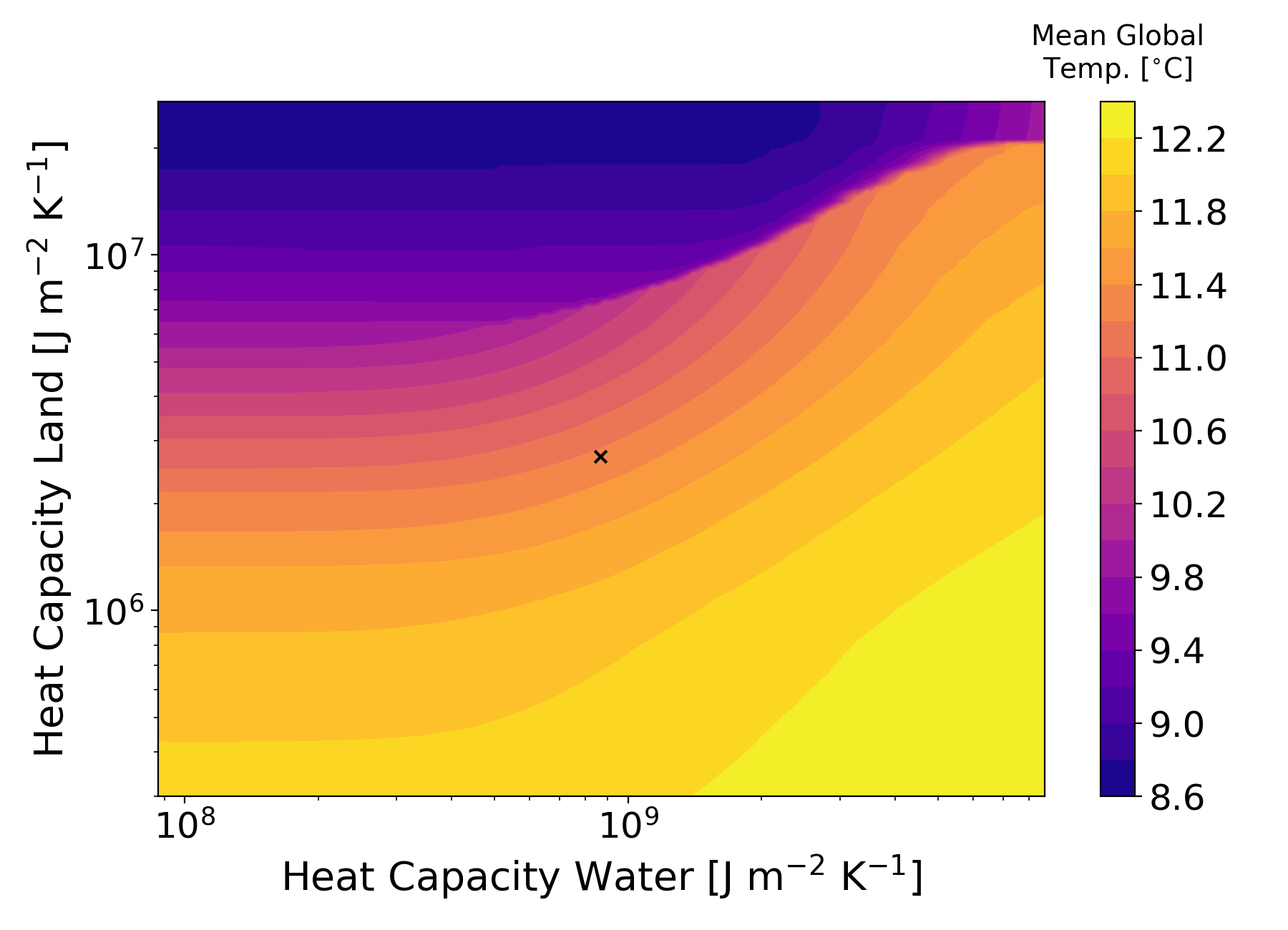}
    \caption{Mean global surface temperature for various heat capacity of land (C$_{L}$) and heat capacity of water (C$_{W}$) configurations. The X marks the calibrated parameters used in this work.}
    \label{fig:heat_cap}
\end{figure*}

Because of this large increase in C$_{W}$, and because the seasonal cycle can be extreme for high eccentricity orbits, we explored the climate's sensitivity to both heat capacities. For this exploration we used $\varepsilon=10^{\circ}$, p\ce{CO2}$=8.49$ bar, and $\psi=0^{\circ}$. We then varied C$_{L}$ and C$_{W}$ by two orders of magnitude each. Figure \ref{fig:heat_cap} shows the resulting mean global surface temperature across this sweep, with the X marking the values adopted in this work. Throughout the full sweep the global mean surface temperature varies only between roughly 8.6 and 12.2$^{\circ}$C, so the climate remains temperate everywhere in this parameter range and never collapses into a snowball state.

The two heat capacities influence the global mean in opposite senses: increasing C$_{W}$ warms the planet, while increasing C$_{L}$ cools it. As expected, for an Earth-like land distribution, higher values of C$_{W}$ allow for more stable climates as the surface is mostly covered by water. Higher water heat capacity values allow for more modest seasonal changes due to the water acting as a buffer for temperature variations. The cooling associated with higher C$_{L}$ reflects the land's lower thermal inertia, which lets it warm quickly at periastron and cool quickly afterward, lowering surface temperatures at higher latitudes, thus favoring ice formation. We conclude that the climate is not strongly sensitive to the precise choice of heat capacities, which gives confidence that small uncertainties in these parameters do not qualitatively change the planet's climate state.

%\textbf{As for Set E, we decided to not vary the precession angle due to the results of Set A which shows how this parameter does not impact the climate. To validate this, we ran 14,000 simulations for modern land distribution at $\psi=[0^{\circ}, 90^{\circ}, 180^{\circ}, 270^{\circ}, 360^{\circ}$], varying the eccentricity and obliquity as shown in Table \ref{tab:simulations}, and fixing p\ce{CO2} to 8.75 bar (as in Figure \ref{fig:landdist}). The results show negligible variation for the final ice states at different precession angles, hence we do not show the results here. However, it does further confirm the low impact precession angle has over the planet's climate.} 

Our simulations assumed that the orbital and rotational parameters do not evolve. Although no other planets are known in the system, an oblate planet will experience rotational axial precession due to the stellar torque \citep{Kinoshita_1975,Kinoshita_1977,laskar1993}. As we have found the precession angle to be relatively unimportant with regard to climate, we expect the role of axial precession to be minimal.

The planet is likely to experience significant tidal effects from its host star \citep{KASTING1993108,barnes2017}, and hence its obliquity, rotational period, eccentricity, and semi-major axis could change. We expect Gl 514 b's rotation rate to tidally damp, but the end state most likely depends on the eccentricity \citep{Hut1981,Correia2008,Barnes_2008,Ferraz2008,Rodriguez2012}. For large values of eccentricity, the rotational frequency is likely to be in a high-order resonance with the orbital frequency such that the planet rotates faster than it orbits. Regardless, the rotational braking will affect its day-night cycle and potentially the atmospheric heat distribution. As the precessional frequency depends on the obliquity and rotation rate,  the rotational axis evolution could be complex and result in an unstable climate evolution. Future research should explore the role of tidal evolution and the stellar torque on this planet.

In this paper, we focused on surface habitability. However, we emphasize that a snowball planet is not necessarily sterile. Life can still exist in subsurface water reservoirs \citep{Lechte2019}, which are plausible as the planet's interior is likely hot enough to maintain them.

%A planet would require a higher instellation to escape from its snowball state \citep{Wilhelm_2022}, unless the planet has processes that can buffer the planet from extreme temperatures such as volcanism and carbonate-silicate weathering \citep{Abbot_2011, Haqq-Misra2016}, creating a metastable climate. Since atmospheric processes are also affected by the interior of the planet, future work should explore the effect of outgassing on the surface temperature. VPLanet contains another module called ThermInt (thermal interior evolution) which in the future can be coupled to the POISE module to study this interaction. As for the occurrence of life on a planet covered in ice, it is hypothesized that a snowball event on Earth triggered evolutionary changes that were made possible by glaciations \citep{Schopf_1992}.} 

In conclusion, we found Gl 514 b to be a potentially habitable world that is worthy of follow-up, both theoretical and observational. Our investigation revealed that the most likely surface conditions for Gl 514 b are ice free or globally ice-covered, with a small chance of ice caps or an ice belt. These states follow from a model whose free parameters we recalibrated against modern Earth's geography, mean surface temperature, OLR, and all four latitudinal ice lines simultaneously. The different climate states could lead to distinct observational signatures at single-pixel resolution, which represents the level of spatial detail expected for future direct-imaging campaigns. Our results and those of \citet{Biasiotti_2024} therefore support more computationally expensive simulations with GCMs to test our results and to generate robust predictions for future telescopes. EBM studies such as these can help streamline these more sophisticated methods to ensure the community is prepared to interpret the data when they arrive. While this investigation cannot definitively evaluate the surface habitability of this planet, it opens the door to future projects that can, and ultimately will, determine whether it is a candidate for biosignature searches with future ground and space-based telescopes.

\section{Acknowledgments}

HEDD and RKB acknowledge support from NASA award No. 80NSSC24K0856. HEDD also acknowledges partial support  from NASA cooperative agreements NNA13AA93A and 80NSSC18K0829. This work was facilitated through the use of advanced computational, storage, and networking infrastructure provided by the Hyak supercomputer system at the University of Washington.

%\appendix

%\section{Appendix information}

\bibliography{climates}{}
\bibliographystyle{aasjournal}

%% This command is needed to show the entire author+affiliation list when
%% the collaboration and author truncation commands are used.  It has to
%% go at the end of the manuscript.
%\allauthors

%% Include this line if you are using the \added, \replaced, \deleted
%% commands to see a summary list of all changes at the end of the article.
%\listofchanges

\end{document}